\pdfoutput=1

\documentclass[11pt]{article}

\usepackage[preprint]{acl}

\usepackage{times}
\usepackage{latexsym}

\usepackage[T1]{fontenc}

\usepackage[utf8]{inputenc}

\usepackage{microtype}

\usepackage{inconsolata}

\usepackage{color}

\usepackage{multirow}
\usepackage{cleveref}
\usepackage{graphicx}

\title{Dynamic Retrieval-Augmented Generation}

\author{Anton Shapkin* \\ JetBrains Research \\ anton.shapkin@jetbrains.com 
        \And  
        Denis Litvinov* \\ JetBrains Research 
        \And 
        Yaroslav Zharov* \\ JetBrains Research 
        \AND 
        Egor Bogomolov \\ JetBrains Research
        \And 
        Timur Galimzyanov \\ JetBrains Research
        \And 
        Timofey Bryksin \\ JetBrains Research \\ timofey.bryksin@jetbrains.com
        }

\begin{document}
\maketitle
\def\thefootnote{*}\footnotetext{These authors contributed equally to this work}
\begin{abstract}
Current state-of-the-art large language models are effective in generating high-quality text and encapsulating a broad spectrum of world knowledge. These models, however, often hallucinate and lack locally relevant factual data. Retrieval-augmented approaches were introduced to overcome these problems and provide more accurate responses. Typically, the retrieved information is simply appended to the main request, restricting the context window size of the model.  
We propose a novel approach for the Dynamic Retrieval-Augmented Generation (DRAG), based on the entity-augmented generation, which injects compressed embeddings of the retrieved entities into the generative model. The proposed pipeline was developed for code-generation tasks, yet can be transferred to some domains of natural language processing. To train the model, we collect and publish a new project-level code generation dataset. We use it for the evaluation along with publicly available datasets. Our approach achieves several targets: (1) lifting the length limitations of the context window, saving on the prompt size; (2) allowing huge expansion of the number of retrieval entities available for the context; (3) alleviating the problem of misspelling or failing to find relevant entity names. This allows the model to beat all baselines (except GPT-3.5) with a strong margin.
\end{abstract}

\section{Introduction}

In the area of natural language and code generation tasks, large-scale pre-trained language models (LLMs) such as T5~\cite{T5}, GPT-3~\cite{GPT3}, and LLaMA~\cite{Llama} have made significant strides.
However, the process of tuning these models is tedious and resource-demanding.
These models have no knowledge on something outside of their training data and are difficult to update.
The scientific community adopted the practice of prompting to fix this problem~\cite{InContextLearning}.
In this paradigm, the prompt includes the relevant information that helps the model to solve the task.
To automate this process of knowledge grounding, the pertinent information is selected from a knowledge base by some similarity measure.
These chunks of information are typically referred to as documents, and the method is referred to as the retrieval-augmented generation (RAG).
For more information on such approaches, we recommend the survey by~\citet{survey}.

RAG methods, however, suffer from some issues.
First, the performance of the generation is limited by the performance of the retrieval model --- if the retrieval model fails to retrieve a relevant document, the generation model will be left with the information available during the training time or even distracted by irrelevant context, providing wrong answers. 
Second, tuning and inference of such models are resource-intensive due to the sheer size of the documents in the knowledge base, typically being prepared for humans to read, not for models to scrape.

To solve these problems, the community proposed a multitude of methods and solutions. 
In order to improve retrieval accuracy, researchers proposed to use the unaugmented model output as a query for the retriever~\cite{GandR} or to expand the query with the additional context generated by a language model~\cite{GAR}. 
To ease the computational burden, the Memorizing Transformers method proposed by~\citet{MemFormer} added a memory bank to the architecture and allowed the generator model to attend to entities from the memory bank, saving computations by caching.

In this work, we target a specific case of RAG that we call \emph{generation with named entities}. For this problem, models can use names associated with the documents during generation. This setup appears in many tasks: for code generation, models can call methods defined in a project~\cite{RepoCoder}, for SQL generation~\cite{Spider} they can use table and column names, for Bash generation~\cite{Nl2Bash} -- command flags. For all the mentioned tasks, knowledge of the available entity names is crucial for generation quality but cannot be captured during the model's initial pre-training. Moreover, the number of available project methods or SQL columns can be tremendous, which makes it infeasible to add them all to the context.

To account for named entities in the code generation task, the work on Repository-Level Prompt Generation (RLPG) by~\citet{RepositoryLevelPromptGeneration} and RepoCoder by~\citet{RepoCoder} proposed specific ways of advanced context gathering.

In this work, we propose DRAG (Dynamic Retrieval Augmented Generation), a novel way to both retrieve named entities and use them during generation. DRAG first encodes the named entities and their contexts (\textit{e.g.,} methods available in the project) to the latent space. Then, it integrates the embeddings into the generator's vocabulary so that during the token-by-token generation, it can also predict an entity name as a single token.
The dynamic nature of retrieval comes from the fact that with DRAG, the generator has access to retrieval data embeddings at each token generation step and is aware of the current context.
An important feature of DRAG is its ability to enhance existing language models, even though by further fine-tuning.
We evaluate our method in Python, SQL, and Bash generation setups. For Python generation, we also collect and publish a large-scale dataset targeting project-level retrieval.

With this work, we make the following contributions:
\begin{itemize}
    \item DRAG, a novel method for generation with named entities that can be used with existing language models to retrieve from a large set of entities on the fly and directly predict entity names during generation.
    \item A large-scale dataset for repository-level code generation that focuses on the evaluation of models' ability to identify and correctly use methods from the project at hand.
    \item Evaluation that demonstrates consistent performance improvements by extending the existing models with DRAG across three different domains on both publicly available datasets and our newly gathered one: Python generation, text-to-SQL task, and Bash generation.
\end{itemize}

\section{Method}

\begin{figure*}[t]
\centering
\includegraphics[width=0.9\textwidth]{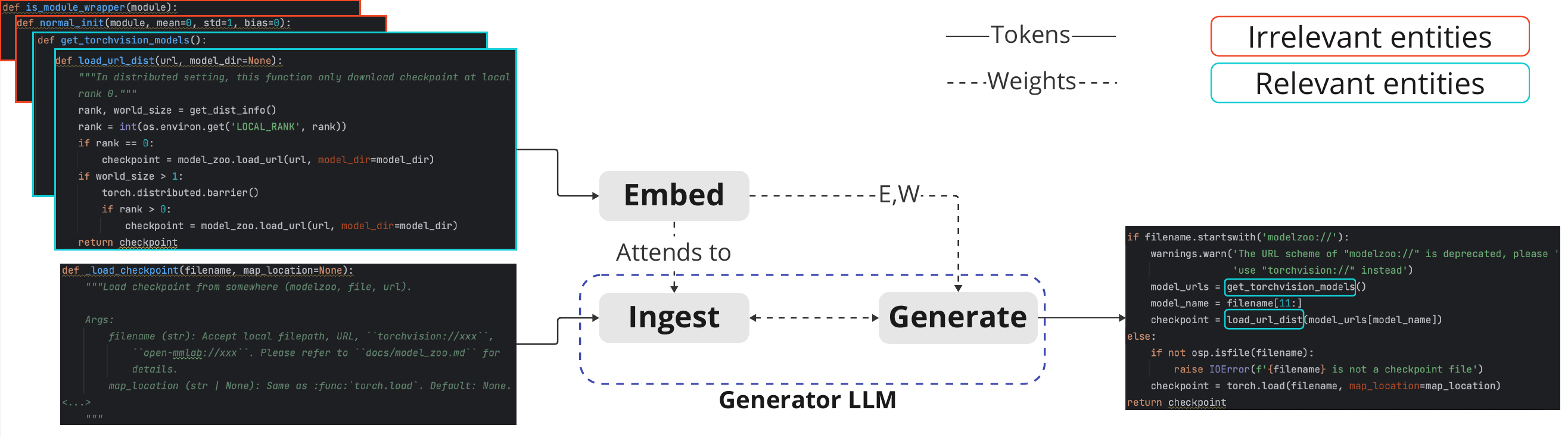} 
\caption{The proposed method architecture. The embedder model produces embeddings of all the documents associated with the sample using cross-attention with the input representation from the generator model. These embeddings extend the generator vocabulary, allowing the generator to decide at each step whether to generate a regular token or to insert an entity.}
\label{architecture}
\end{figure*}

To define our method more precisely, we start with the notation.
We denote one separate chunk of context as a document if it represents a description of a single named entity.
For the sake of simplicity, we consider cases where extracting the entity name is trivial.
For example, in the case of the code generation task, one document will be the function declaration, and the corresponding entity name will be the function name.

To improve the  entity-augmented generation, we propose to reimagine the stages of the typical retrieval-augmented generation pipeline.
Typically, the retrieval is first performed by a separate model, whether sparse, \textit{e.g.}, BM25~\cite{BM25}, or dense, \textit{e.g.}, DPR~\cite{Karpukhin2020}.
Then, the retrieved documents are fed into the generation model, either as an embedded representation, \textit{e.g.}, in Memorizing Transformers~\cite{MemFormer}, or as a sequence of tokens, \textit{e.g.}, in RepoCoder~\cite{RepoCoder}.

Following the findings by~\citet{EmbeddingsReveal}, which show how much information is packed inside the embeddings provided by LLMs, we hypothesize that modern LLMs may not need the whole text of entities to gain useful information from the document.
On the other hand, we note that there is an entire class of tasks where the retrieved documents coincide with the entities that models need to use during generation. 
For example, the Question-Answering task requires extracting information from the documents, but in the API call prediction in coding tasks, one type of the used documents is function code, while the entity to be used during generation is the function name.
Current approaches, as discussed, rely on a retriever model to find relevant documents and then on a generator model to extract and use the entity name from the document.

We redefine the components of the retrieval-augmented generation pipeline as the \emph{embedder} and \emph{generator} models as shown in~\Cref{architecture}.
The proposed method goes through the following steps to generate one sample.
First, all documents that can theoretically be used for the generation are considered available, \emph{e.g.}, for the code generation --- all code in the same repository will be considered available.
The embedder compresses representations of all available documents into a set of embeddings. These embeddings are added (after some transformation) to the generator model's vocabulary as a single token for each entity name.
The generator then performs the prediction in a usual token-by-token manner, while expanded vocabulary allows predicting the embedded representation of a document instead of a regular token.
During the decoding process, the predicted embeddings are replaced with entity names directly extracted from the documents via extended vocabulary.

The change in the architecture serves several purposes.
On the one hand, by incorporating the documents as the extension of the generator's vocabulary instead of the textual context in the prompt, we lift the length limitations of the context window.
This, in turn, greatly expands the possible number of documents available for the model, given a good embedder model. 
On the other hand, by allowing the generator to refer to entities as a whole instead of generating their name token-by-token, we alleviate the problem of misspelling the entity name when copying~\cite{SolidGoldMagikarp} or failing to find them in the context if it is too long~\cite{NeedleInaHaystack}.

The common thread among the recent works in this area is leveraging pre-trained LLMs, without fine-tuning them.
However, state-of-the-art LLMs require significant computational resources to run. We focus on small and medium-size models (300M--2B parameters), which we fine-tune for retrieval tasks.
For that, we collect and publish a new large-scale dataset for project-level code generation.

The rest of this section describes particular elements of the proposed architecture.
First, we describe the embedder and the generator models.
Then, we address the training process.

\subsection{Embedder}

For this work, we consider dense embeddings provided by Transformer-based models. This way, we can keep the embedder consistent with the generator model.
In particular, we consider tuning encoders from T5 (60M)~\cite{T5} and CodeT5 (60M)~\cite{CodeT5} for the architecture simplicity and demonstrated performance.
In this work, the embedder model is also referred to as $f_e$.

To improve the document summarization and make it dependent on the input, we experiment with adding cross-attention between the embedder and the generator part responsible for the input ingestion, 
$$softmax \left(\frac {Q_g^T K_e} {\sqrt d}\right)V_e, $$
where $Q_g$ is the query matrix from the last layer of the generator model $g$, $K_e$ and $V_e$ are key and value matrices of the retriever $e$, and $d$ is the hidden size.
If the generator has an encoder-decoder architecture, we use the encoder. If the generator has the decoder-only architecture, we use a slice of the query matrix $Q_{g;1:T}$, where $T$ is the length of the input in tokens.
In~\Cref{architecture}, this part of the generator is labeled as ``Ingest''.

If the inference efficiency is a larger concern than the quality, our method can offer two possibilities.
The first one is to follow the Memorizing Transformers~\cite{MemFormer} and cache the $K_e$ and $V_e$ matrices.
The second possibility is to remove the cross-attention between the embedder and the generator and cache the embeddings for the entire set of documents.

\subsection{Generator}
In this work, we consider the generator to be a Transformer with an encoder-decoder or decoder-only architecture.
In the experiments, we use T5 (738M)~\cite{T5}, OpenLLaMA (3B) ~\cite{openlm2023openllama}, and CodeGen (350M, 2B)~\cite{CodeGen} models for this purpose.
We will refer to this model as $f_g$.

To allow the generator to select an entity from the set of available entities during generation, we extend the model's vocabulary with the entity embeddings provided by the embedder model.
For a given input, we perform the vocabulary extension in the following way.
First, all available documents $\{\mathcal{D}_i\}_{i=0}^M$ are converted into a set of embeddings $\mathcal{E}_i = f_e(\mathcal{D}_i)$.
Next, we use two separate Multilayer Perceptrons (MLPs) $f_\varepsilon, f_w$ to convert those embeddings into the columns of the generator's embedding matrix $E'_i = f_\varepsilon(\mathcal{E}_i)$ and the output linear layer $W'_i = f_w(\mathcal{E}_i)$, respectively.
The input embeddings $E$ and output linear layer $W$ of the generator model $f_g$ are then replaced by their extended versions $E \mapsto E \Vert E'$ and $W \mapsto W \Vert W'$, where $\Vert$ denotes concatenation.
After this set of operations, we can treat the generator model as if we just extended its vocabulary with a set of special tokens. Note that the vocabulary is re-extended for each sample.

\subsection{Training and Decoding}
We propose two possible modes of the model training.
The first is an end-to-end training of both the generator and embedder.
The second is training only the generator, together with the MLPs $f_\varepsilon, f_w$.
In both cases, we use the classical log-likelihood loss for the task of next token prediction, used for the language modeling.

We hypothesize that if the generator and the embedder are initialized with the same weights, there could be a way to fine-tune only the generator.
Should this prove to be the case, apart from a usual generator model fine-tuning, we need to tune only the MLP connectors, converting the representations of the entities from the embedder's latent space to the space of the generative model's token embeddings. We test the hypothesis in \Cref{sec:ablation} and for other experiments stick to training both the generator and embedder.

\section{Related Work} \label{related_work}

Like the R2-D2 model by~\citet{R2-D2}, who composed a pipeline of a retriever, reranker, extractive reader, generative reader, and an aggregation mechanism, we aim to shorten the documents and provide the model with easier access to the entities.
However, in contrast to them, we do not use the sequence-of-tokens representation at all, and we aim to solve the generation task.

Matching the PICARD model proposed by~\citet{PICARD}, which uses constrained beam search for execution-guided decoding to enforce syntax correctness, we interfere with the model's generation procedure, forcing it to produce correct entities.
Yet, unlike this method, we do not change the sampling procedure but allow the model to select desired entities in the same way it selects usual tokens.
The same difference lies between our work and ReCode~\cite{RECODE}.

Similar to the Memorizing Transformers approach~\cite{MemFormer}, which stores the documents in a large cached database, we transform the documents into their representations before adding them to the model.
However, we do form the document set dynamically and do not explicitly retrieve the relevant documents as a separate step.

Finally, like REALM~\cite{REALM}, RetGen~\cite{RetGen}, RAT-SQL~\cite{RAT-SQL}, ReACC~\cite{ReACC}, RepoCoder~\cite{RepoCoder}, CoCoMIC~\cite{ding2023cocomic} and many others, discussed in the survey by~\citet{survey}, we work on the approach of retrieval-augmented code generation.
However, we combine it with entity-augmented generation and also combine the retriever and the generator in one model, relaxing some of the persistent constraints.

We would especially like to stress our connection with the RLPG and RepoCoder~\cite{RepositoryLevelPromptGeneration,RepoCoder}. 
RLPG proposed a heuristic set of rules to select a relevant context from the current repository and a model to predict which set of rules will work best for a specific case.
RepoCoder iteratively alternates two stages: (i) retrieval based on the model prediction and (ii) prediction based on the retrieved information.
RepoCoder is the standing SoTA for repository-level code generation. 
Therefore, we use it for experiments as a smart way to retrieve the context. 
However, contrary to RepoCoder, we do not modify the initial entity selection step and focus on how the model ingests the context.

We acknowledge the concurrent method ToolkenGPT by~\citet{TookenGPT}, who introduced tunable tokens for the tool-augmented LLMs.
However, our key difference is the dynamic nature of the vocabulary extension proposed in our work.

\section{Experiments}
\label{sec:experiments}
We evaluate the performance of different approaches across three datasets where the result needs to be generated using entities from the corresponding knowledge base: 
the novel dataset for project-level code generation, the NL2Bash dataset~\cite{Nl2Bash}, and the Spider SQL dataset~\cite{Spider}. 
For each dataset, we report task-specific metrics along with the \emph{recall} for the generator's entity prediction. Recall is computed as a ratio of the entity names in the ground truth solution that were also predicted by the model.

The rest of this section is organized as follows.
First, we discuss the baselines we will use to compare our model to.
Then, for each task, we briefly describe the dataset, metrics, baselines, and 
 results.

 \subsection{Baselines}

We follow RepoCoder~\cite{RepoCoder} and RLPG~\cite{RepositoryLevelPromptGeneration} when selecting baselines.
We consider each method from two points of view: how it filters the context and how it utilizes the filtered context.
For reference, the RepoCoder method provides a novel perspective on context filtering, retrieving the APIs from the repository in a multi-step procedure. 
However, it uses the classical way of context utilization, appending the text of the retrieved snippets to the prompt.

\paragraph{Context filtering methods}

\paragraph{No} designates the case where the context is completely absent. 
    The only knowledge available to the generation model is stored in its weights and request.

\paragraph{All} designates the case where the context comprises all possible knowledge, \textit{e.g.}, the entire code repository or a knowledge base.

\paragraph{k-NN} stands for the k Nearest Neighbours algorithm run on the distances in the embedded space.
    The relevant context is found by selecting 5 documents from the knowledge base that have the closest embeddings to the input.
    To produce the embeddings, we used BERT sentence embeddings~\cite{BERT}.

\paragraph{Oracle} describes the context containing only documents to be used in the sample.
    This baseline tests the model's ability to utilize the provided context, given its highest quality.

\paragraph{RepoCoder} ~\cite{RepoCoder} proposed a novel way of the multi-step context collection.
    First, the unfinished code is utilized for the initial retrieval.
    Then, the sparse retrieval is used to obtain code snippets that are relevant to the current prediction from the whole repository.
    The retrieved snippets are fed into the generator model to complete the function body.
    After this initial step, an alternation of retrieval based on previous prediction and generation based on the new retrieval is used to improve the quality.

\paragraph{Context utilization methods}

\paragraph{Prompt} describes the situation when the generator model is given access to the context by appending it in textual form to the input query.
This is the leading way to give additional context to the generator model, adopted by~\citet{RepoCoder,RetGen,RepositoryLevelPromptGeneration} and others.

\paragraph{DRAG} is the proposed method of the context utilization.
We extend the model's vocabulary with special tokens representing the entities from the knowledge base.
The model can then utilize these entities during the generation.

\subsection{Project-level code generation task}

\paragraph{Dataset.} The dataset (see~\Cref{sec:dataset} for details on the dataset collection) consists of around 16,000 samples derived from around 1,000 GitHub repositories in Python. 
In this task, a model should generate a function body given the function signature and the docstring while using a set of functions implemented in the project. 
Functions available in the repository comprise a set of entities. We take function names as entity names and the whole function as the entity content. 
We follow the work by~\citet{OutoftheBLEU} and utilize the ChrF~\cite{chrf} as the dataset-specific metric to assess code quality.

\paragraph{Models and Implementations.} 
We implement this experiment's baselines and DRAG with the CodeGen-Mono-350M and CodeGen-Mono-2B models.
Additionally, we report the results of the RepoCoder and \textit{Oracle} context filtering with the GPT-3.5 model.
The embedder model for the DRAG method was T5-small and was tuned together with the generator (w/o cross-attention).

\begin{table}[t]
\centering
\begin{tabular}{llcc}
\hline
\textbf{Context} & \textbf{Utilization} & \textbf{chrF} & \textbf{recall} \\
\hline
\multicolumn{4}{c}{\texttt{CodeGen-Mono-350M}}\\\hline
No & Prompt & 23.1 & 0.07 \\
RepoCoder & Prompt & 25.2 & 0.27 \\
Oracle & Prompt & 27.9 & 0.22 \\
All & \textbf{DRAG} & 38.5 & 0.35 \\
RepoCoder & \textbf{DRAG} & 40.0 & 0.48 \\
Oracle & \textbf{DRAG} & 42.1 & 0.74 \\
\hline
\multicolumn{4}{c}{\texttt{CodeGen-Mono-2B}}\\
\hline
No & Prompt & 33.7 & 0.14\\
RepoCoder & Prompt & 29.1 & 0.32 \\
\hline
\multicolumn{4}{c}{\texttt{GPT-3.5}}\\
\hline
RepoCoder & Prompt & 42.8 & 0.39 \\
Oracle & Prompt & 42.5 & 0.93\\\hline
\end{tabular}
\caption{Performance study of the proposed approach on the newly collected repository-level code generation dataset. The task is code generation given the function signature.}
\label{performance-our-dataset}
\end{table}

\paragraph{Results Discussion.} We report the collected metrics in~\Cref{performance-our-dataset}.
We draw three conclusions here: \textbf{(i)} DRAG consistently outperforms prompting methods given the same filtering and model size, \textbf{(ii)} DRAG enables feeding the model with all available entities at once,  \textbf{(iii)} being supported with a good filtering strategy, a small model equipped with DRAG outperforms larger models without it.

\subsection{NL2Bash task}
\label{nl2bash}
\paragraph{Dataset.} The dataset ~\cite{DBLP:journals/corr/abs-2103-02523} consists of around 8,000 training samples and 600 samples for validation and testing. In this task, a model should generate Bash commands based on a natural language description, given the \texttt{man} page documentation. File names and the order of flags are not taken into account during evaluation.  

Bash commands can form a chain or have nested commands. The distribution of utility usage across samples is far from uniform, with just several utilities adding up to 60\% of samples. For this reason, the main challenge is to predict utility flags correctly. In this dataset, we use flag names as entity names and utility flag documentation as entity descriptions.

To measure the quality, we follow the advice of the dataset authors and measure the Exact Match (EM) together with the \emph{recall} for the generator's entity prediction.

\paragraph{Models and Implementation.} For this experiment, we implement the \textit{No}-context and \textit{k-NN} filtering baselines with Prompt utilization strategy and compare it with \textit{All}-entities and \textit{Oracle} filtering with DRAG.
The model architecture is LLaMa-3B, with T5-small used as the embedder model for DRAG (with cross-attention).
Additionally, as a reference point of the accessible results, we report the GPT-3.5 results prompted with all flags available for the relevant commands.

\begin{table}[t]
\centering
\begin{tabular}{llcc}
\hline
\textbf{Context} & \textbf{Utilization} & \textbf{EM} & \textbf{recall} \\
\hline
\multicolumn{4}{c}{\texttt{LLaMa-3B}}\\\hline
No & Prompt & 7.1 & 0.06\\
k-NN & Prompt & 12.2 & 0.34\\
All & \textbf{DRAG} & 12.8 & 0.36\\
Oracle & \textbf{DRAG} & 55.2 & 0.75\\\hline
\multicolumn{4}{c}{\texttt{GPT-3.5}}\\\hline
All & Prompt & 28.4 & 0.60\\\hline
\end{tabular}
\caption{Performance study on the NL2Bash dataset by~\citet{Nl2Bash}. The model used for this experiment is LLaMa-3B for the generator and T5-small for the embedder.}
\label{performance-nl2bash-dataset}
\end{table}

\paragraph{Results Discussion.} From~\Cref{performance-nl2bash-dataset}, we see that the proposed model outperforms all the baselines apart from the \textit{Oracle}.
We hypothesize that the small margin could be caused by the fact that the grammar allows combining flags and having several versions of the same flags.
For example, the \verb|--verbose -a| flag can be replaced by \verb|-Va|.
The inability to modify the entities, even when the model can learn the rules of such modification, is a notable disadvantage of the DRAG method.
As an argument towards this hypothesis, we point out the performance of the \verb|GPT-3.5| model prompted with all relevant flags.

\subsection{Spider SQL task}
\paragraph{Dataset.} The dataset~\cite{Spider} consists of around 7,000 samples in the train set and 1,000 samples in the dev set. In this task, a model should generate an SQL query from a natural language input, given the database schema. 
Each sample from the dataset refers to its own database.

We use a popular data preprocessing technique from the Spider leaderboard, simplifying SQL queries using NatSQL~\cite{NatSQL}. 
The NatSQL simplification essentially makes outer join on the tables in the database, providing table aliases to the columns in the \texttt{table.column} form. 
We concatenate the column name with the corresponding values from the database records to form documents while using the \texttt{table.column} as entity name.
The model trains to generate NatSQL queries, which are subsequently converted to the full SQL form. The dataset metrics are evaluated on the full SQL form.

We follow the dataset authors in using the Exact Match (EM) and Execution Accuracy (EX)  to measure model performance. EM evaluates how much the generated SQL is comparable to the ground truth, while EX measures how well it matches the output of the execution of the generated SQL.

\paragraph{Models and Implementation.} For this experiment, we used the T5 generator model, with T5-small as the embedder model (with cross-attention).
Same as in~\Cref{nl2bash}, we implement the \textit{No}-context and \textit{k-NN} filtering baselines with Prompt utilization strategy and compare it with All-entities and \textit{Oracle} filtering with DRAG.
As a reference point, we report the results of GPT-3.5 prompted with all column names of the database.

\begin{table}[t]
\centering
\begin{tabular}{llcc}
\hline
\textbf{Context} & \textbf{Utilization} & \textbf{EX} & \textbf{recall} \\
\hline
\multicolumn{4}{c}{\texttt{T5}}\\\hline
No & Prompt & 16.3 & 0.32\\
k-NN & Prompt & 41.9 & 0.63\\
All & \textbf{DRAG} & 63.9 & 0.78\\
Oracle & \textbf{DRAG} & 80.8 & 0.97\\\hline
\multicolumn{4}{c}{\texttt{GPT-3.5}}\\\hline
All & Prompt & 74.1 & 0.86\\\hline
\end{tabular}
\caption{Performance study of the model on the Spider dataset by~\citet{Spider}. The task is text-to-SQL. The models used for the test are T5 as the generator and T5-small as the embedder.}
\label{performance-spider-dataset}
\end{table}

\paragraph{Results Discussion.} As we demonstrate in~\Cref{performance-spider-dataset}, the proposed method outperforms all baselines but the \emph{Oracle} by a solid margin.
The high score of the approach using all the available context together with the DRAG suggests that for cases with a high diversity of entities and the density of their appearance, the entity filtering step can be excluded altogether.

\subsection{Practitioner Questions}\label{sec:ablation}

Finally, we do an ablation study to evaluate parts of the DRAG method proposed in this work. 
We evaluate whether the entity embedder should be trained in order to improve the overall quality. 
Then, we compare our method of retrieving entity names from vocabulary to the conventional retrieval from the prompt. 
We conduct both experiments on the newly proposed dataset for code generation described in detail in \Cref{sec:dataset}.

\paragraph{Is embedder training required?}

\begin{table}[t]
\centering
\begin{tabular}{lccc}
\hline
\textbf{Parts Tuned} & \textbf{chrF} & \textbf{recall} & \textbf{precision} \\\hline
\verb|E+G| & 38.5 & 0.35 & 0.27\\
\verb|G| & 39.0 & 0.34 & 0.30\\\hline
\end{tabular}
\caption{Ablation study on the joint training of the embedder (E) and the generator (G) models.}
\label{ablation-generator-only}
\end{table}

We hypothesize that given the adaptive abilities of the generative models, the joint training of the embedder and the generator models may not be needed.
To check this, we additionally tuned one instance of DRAG without optimizing the embedder.
The results are presented in~\Cref{ablation-generator-only}.
We conclude that training of the embedder model is not a strict requirement.

\paragraph{Is retrieving from the vocabulary beneficial compared to prompting?}

Our method introduced the novel entity retrieval approach, where the model can select an entity from the vocabulary instead of copying the entity name from the prompt.
The performance experiments presented in~\Cref{sec:experiments} demonstrate that this is generally beneficial.
To add to this comparison, we note that (i) DRAG outperforms Prompt when both are fed with the \textit{Oracle} entities, and (ii) when fitting the full context in smaller models is impossible, DRAG is not only able to fit the full context in, but makes a step in quality towards the \verb|GPT-3.5| level.

\section{DRAG Code Generation Dataset}\label{sec:dataset}

Existing datasets for code generation either do not have a repository-level context or are not tailored towards training or fine-tuning language models, comprising small sets of examples. For instance, the dataset collected by RLPG~\cite{RepositoryLevelPromptGeneration} consists of 47 repositories.
While sufficient to train a small classifier model, as proposed in their method, it will not satisfy the requirements for the generator model fine-tuning of our approach.
The same appears to be true for the datasets from RepoCoder~\cite{RepoCoder} and PragmaticCode introduced by~\citet{MGD}, with 14 and 100 repositories, respectively.

Hence, to train our models for repository-level code generation, we collected a \textbf{repository-level code generation} dataset of around 16,000 samples derived from around 1,000 GitHub repositories in Python, divided into train/dev/test sets in the ratio of 80/10/10 without an overlap of repositories between the sets.
Each sample includes a method signature, the repository-wide context given as input, and the respective method body as target output.

We considered the following criteria for data mining to make baseline evaluation feasible. Firstly, the repository should be created later than 01.01.2022 to avoid data leakage, and have at least 10 stars and 5 watchers to select good-quality code~\cite{pitfalls}. Secondly, the target function with intra-project calls should (a) contain from 30 to 400 tokens to reduce the hallucination effect during sequence generation and (b) have a docstring.
In addition, we paid careful attention to the licensing, and selected only repositories that have one of the following licenses: Apache-2.0, MIT, BSD-3-Clause, and BSD-3-Clause-Clear, which are the most prevalent permissive licenses for Python projects. This ensures that our dataset can be used for further studies.

On average, the collected repositories have 55 functions in the context (median 60), the target function has a length of 70 tokens (median 60), and there are 1.45 calls of project functions (median 1) per sample.
We will publish the dataset alongside the paper upon acceptance. We attach a sample of data and source code to this submission. 

To assess the models' quality on our dataset, we propose to focus on the quality of code generation as well as the entity retrieval quality, thus measuring how well the entity was retrieved and then utilized for the generation. 
We propose to use ChrF~\cite{chrf} to assess the quality of code generation following the study on metrics' quality by~\citet{OutoftheBLEU}. 
To assess the entity retrieval quality, we suggest computing recall and precision of entity prediction, comparing entity names in the reference solution and the ones predicted by the generator (without accounting for their positions).

\section{Limitations}
We consider the following limitations of our work.
    
The size of the models used for comparison is relatively low. 
While we consider our research substantial for small-scale models, larger models are lacking in this paper.
We argue that the models of larger size are not likely to demonstrate worse results, and we conduct a comparison with baselines of larger size.

The predictions of the DRAG can not modify the predicted entities to make them match the context in the grammatical sense.
This either requires a multi-pass approach or changing the model to accommodate flexibility. However, we argue that in some tasks---including code generation---such modifications are not needed.

\section{Conclusion}

In this paper, we proposed a novel pipeline for the dynamic retrieval-augmented generation (DRAG).
We evaluated it on publicly available datasets, comparing it to both widely used baselines and state-of-the-art methods.  

The new architecture achieves several targets: (1) increasing the length limitations of the context window, saving on the prompt size; (2) allowing a huge expansion of the number of retrieval documents available for the context; (3) alleviating the problem of misspelling or failing to find entity name when copying them from context. This allows the proposed approach to improve the results twofold compared to the baseline prompting.

For our experiments with repository-level code generation, we collected a novel large-scale repository-level code generation dataset.
We will publish the dataset and the code to reproduce our method upon acceptance. We attach a sample of data and source code to this submission. 

While we test our method in the Python, SQL, and Bash generation settings, we hypothesize that it can be beneficial in other generation cases where the model is required to use the pre-defined entities, \textit{e.g.}, in legal or medical texts, documentation and question-answering systems.

\section*{Computational Resources and Libraries Used}
We implement our code in PyTorch and use Transformer implementations from the HuggingFace library~\cite{HuggingFace}.
We also take pre-trained model weights from the HuggingFace hub. 
We conduct all experiments on eight NVIDIA A10 and one A100 GPUs using Ubuntu 22.04 LTS with 256 GB RAM.
In all experiments, we optimize model weights with the Adam optimizer~\cite{Adam} with \texttt{1e-4} learning rate and cosine learning rate schedule~\cite{cosine} for 20 epochs. 
The generation is performed with beam search decoding with a beam size of 5.

\bibliography{custom,references_Yaroslav}

\end{document}